\documentclass[twocolumn,preprintnumbers, amssymb,amsmath,aps,floatfix,prd,nofootinbib,superscriptaddress,showpacs]{revtex4-1}
\usepackage{epsfig}
\usepackage{bm}
\usepackage{amssymb}
\usepackage{amsmath}
\usepackage{color}
\usepackage{subfigure}
\usepackage[colorlinks,
            linkcolor=blue,
            anchorcolor=black,
            citecolor=blue
            ]{hyperref}

\newcommand{\beq}{\begin{eqnarray}}
\newcommand{\eeq}{\end{eqnarray}}

\newcommand{\nn}{\nonumber \\}

\begin{document}
\title{Angular dependence in transverse momentum dependent diffractive parton distributions at small-$x$}

\author{Yoshitaka Hatta}
\affiliation{Physics Department, Building 510A, Brookhaven National Laboratory, Upton, NY 11973, USA}
\affiliation{RIKEN BNL Research Center,  Brookhaven National Laboratory, Upton, New York 11973, USA}

\author{Feng Yuan}
\affiliation{Nuclear Science Division, Lawrence Berkeley National
Laboratory, Berkeley, CA 94720, USA}

%\date{\today}

\begin{abstract}
We discuss the angular dependence of the recently proposed transverse momentum dependent quark and gluon  diffractive parton distributions at small-$x$. We introduce the difractive versions of the Sivers function and the elliptic gluon Wigner distribution and   evaluate them in simple models with gluon saturation and study their geometric scaling properties.  We also show that the diffractive version of the linearly polarized gluon distribution identically vanishes to leading order. These distributions enrich the physics opportunities for measuring semi-inclsuive diffractive DIS  processes at the future electron-ion colliders.
\end{abstract}
%\pacs{24.85.+p, 12.38.Bx, 12.39.St, 12.38.Cy}
\maketitle

\section{Introduction}

\begin{figure}[t]
\begin{center}
\includegraphics[width=0.3\textwidth]{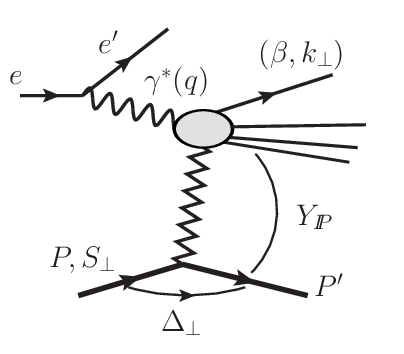}
\end{center} 
\caption[*]{ A schematic picture of semi-inclusive diffractive DIS (SIDDIS). Due to a color singlet (`Pomeron') exchange, the elastically scattered nucleon and the  diffractively produced system are separated by a rapidity gap $Y_{I\!\!P}=\ln 1/x_{I\!\!P}$. The produced system itself has a rapidity span $Y_\beta =\ln 1/\beta$ and $x_B=\beta x_{I\!\!P}$ is the usual Bjorken variable.   }
\label{fig:siddis}
\end{figure}

Diffractive  Deep Inelastic Scattering (DIS) has  long been argued to be one of the most promising channels to probe the gluon saturation in hadrons and nuclei at small-$x$ \cite{Bartels:1996ne,Wusthoff:1997fz,Buchmuller:1998jv,Golec-Biernat:1999qor,Kovchegov:1999ji,Kowalski:2006hc,Hatta:2006hs,Marquet:2007nf,Kowalski:2008sa}. The process is characterized by an elastically scattered target separated by a large rapidity gap $Y_{I\!\!P}=\ln 1/x_{I\!\!P}$ from the fragments of the virtual photon (see Fig.~\ref{fig:siddis}). Typically, the diffractively produced system consists of a single vector meson \cite{Golec-Biernat:1999qor,Kowalski:2006hc} or 2-3 jets \cite{Bartels:1996ne,Altinoluk:2015dpi,Hatta:2016dxp,Boussarie:2016ogo,Boussarie:2019ero,Fucilla:2022wcg,Zhou:2016rnt,Hagiwara:2017fye,Mantysaari:2019csc,Mantysaari:2019hkq,Iancu:2021rup,Iancu:2022lcw,Hauksson:2024bvv} (exclusive diffraction), or it can also be any hadronic final states which span a rapidity range $Y_\beta=\ln 1/\beta$ (inclusive diffraction). Due to the color singlet exchange between the virtual photon and the target, at high energy the cross section behaves like the square of the gluon distribution, resulting in a stronger sensitivity to the gluon dynamics in the nonlinear regime than in non-diffractive processes. 

Recently, the notion of semi-inclusive diffractive DIS (SIDDIS) has been proposed in \cite{Hatta:2022lzj}. This is literally a hybrid of diffractive DIS and semi-inclusive DIS (SIDIS). Namely, one requires the elastically  scattered target and a rapidity gap $Y_{I\!\!P}$  as in inclusive diffraction, but at the same time detects  a single species of hadrons out of the diffractively produced system. Compared to SIDIS, naturally SIDDIS is  more sensitive to the gluon saturation. Compared to diffractive DIS, SIDDIS has an additional kinematical handle, the transverse momentum $k_\perp$ of the observed hadron, which brings in a number of interesting new features as is   familiar in the context of SIDIS. In a sense, SIDDIS is similar to diffractive dijet production  where the jet transverse momentum plays the role of $k_\perp$, and indeed their cross sections are closely related \cite{Hatta:2016dxp}.   From  experimental point of view, an advantage of SIDDIS is that one does not have to reconstruct jets in the final state which may be an increasingly  challenging task as $|k_\perp|$ becomes smaller.  On the other hand, it  requires the measurement of the invariant mass $M_X^2=\frac{1-\beta}{\beta}Q^2$  as well as particle identification of the produced system. 

The basic ingredient in describing SIDDIS is the transverse momentum dependent (TMD) diffractive parton distribution functions (DPDFs) \cite{Hatta:2022lzj}. This is the TMD version of the standard collinear DPDFs \cite{Berera:1995fj}  and can be in principle rigorously defined once QCD factorization is established. At small-$x$, they can be analyzed within the frameworks of high energy factorization and the Color Glass Condensate (CGC)~\cite{Mueller:1993rr,Mueller:1999wm,McLerran:1993ni,McLerran:1993ka,McLerran:1994vd}.     
In Ref.~\cite{Hatta:2022lzj}, the quark and gluon TMD DPDFs at small-$x$ have been evaluated directly from their operator definitions  in terms of the color dipole S-matrix, the ubiquitous building block of parton distributions and QCD amplitudes in the saturation regime. The results are consistent, in appropriate kinematical limits, with the earlier calculations of the diffractive structure functions \cite{Wusthoff:1997fz,Buchmuller:1998jv,Hautmann:1998xn,Golec-Biernat:1999qor,Hautmann:2000pw} recently extended to next-to-leading order \cite{Beuf:2024msh} in the CGC framework.  In the forward limit, the gluon TMD DPDF is  equivalent to the `Pomeron unintegrated gluon distribution' studied in \cite{Iancu:2021rup,Iancu:2022lcw,Hauksson:2024bvv}.

In this paper, we investigate the azimuthal angle  correlation in the quark and gluon TMD DPDFs. In addition to $k_\perp$, TMD DPDFs depend on another transverse vector  $\Delta_\perp$, the nucleon recoil transverse momentum. Moreover, if the incoming nucleon is transversely polarized, there is a third transverse vector $S_\perp$, the nucleon spin vector. One would then expect that these vectors strongly  correlate with each other and generate a rich pattern of  angular dependencies within the distributions. The situation is actually similar to the quark and gluon Wigner distributions which also depend on these vectors. This is so because both the Wigner distributions and TMD DPDFs are related to the dipole S-matrix such that, very roughly, $({\rm TMD\ DPDF})\sim ({\rm Wigner})^2$. There are however interesting differences, as we shall explore below. 

A complementary study can also be carried out for the  moderate/large-$x$ kinematics, where the above-mentioned correlations  provide access to the quark/gluon Wigner distributions in the respective kinematics. A more general framework to address this physics is the so-called fracture functions or target fragmentation functions~\cite{Trentadue:1993ka,Berera:1995fj,Grazzini:1997ih,Graudenz:1994dq,deFlorian:1995fd,Collins:1997sr}. Recent developments along this direction have also revealed opportunities to explore the relevant aspects of nucleon tomography~\cite{Anselmino:2011ss,Anselmino:2011bb,Boglione:2016bph,Chen:2021vby,Chen:2023wsi,Chen:2024brp,Guo:2023uis} in measurements at JLab~\cite{CLAS:2022sqt} and potential measurements in future experiments at JLab and the electron-ion collider (EIC)~\cite{Boer:2011fh, AbelleiraFernandez:2012cc, Accardi:2012qut,AbdulKhalek:2021gbh,Gross:2022hyw,Achenbach:2023pba}.

\section{Diffractive PDFs from Dipole Amplitude at Small-$x$: General Framework}

The definition of the TMD DPDFs has been given in Ref.~\cite{Hatta:2022lzj} which are reproduced here for  completeness. For the quark TMD DPDF, we have 
\begin{eqnarray}
&&2E_{P'}\frac{df_q^{D}(x,k_\perp;x_{I\!\!P},t)}{d^3P'}=\int
        \frac{d\xi^-d^2\xi_\perp}{2(2\pi)^6}e^{-ix\xi^-P^++i\vec{\xi}_\perp\cdot
        \vec{k}_\perp} \nonumber\\
        &&~\times \langle
PS|\overline\psi(\xi){\cal L}_{n}^\dagger(\xi)\gamma^+|P'X\rangle \langle P'X|{\cal L}_{n}(0)
        \psi(0)|PS\rangle  \ ,\label{tmdun}
\end{eqnarray}
with a future pointing gauge link in the fundamental representation ${\cal L}_{n}(\xi) \equiv \exp\left(-ig\int^{\infty}_0 d\lambda \, v\cdot A(\lambda n +\xi)\right)$. In the above definition, we have chosen that the nucleon is moving along $+\hat z$ direction that its momentum $P$ is dominated by its plus component. A light-cone vector $n$ is conjugate to the nucleon momentum $n^2=0$ and $n\cdot P=1$. The momentum transfer between the initial and final states is denoted by $\vec{\Delta}=P'-P$ and $t=\Delta^2$. Furthermore, we define $x_{I\!\!P}= n\cdot (P-P')$ to represent the momentum fraction carried by the Pomeron and $\beta=x/x_{I\!\!P}$. 
Similarly, we can define the gluon TMD DPDF
\begin{eqnarray}
&&2E_{P'}\frac{df_g^{D}(x,k_\perp;x_{I\!\!P},t)}{d^3P'}=\int
        \frac{d\xi^-d^2\xi_\perp}{xP^+(2\pi)^6}e^{-ix\xi^-P^++i\vec{\xi}_\perp\cdot
        \vec{k}_\perp} \nonumber\\
        &&~\times\!\! \langle
PS|F^{+\mu}(\xi){\cal L}_{n}^\dagger(\xi)\gamma^+|P'X\rangle \!\langle P'X|{\cal L}_{n}(0)
        F_\mu^{\, +}(0)|PS\rangle  \ .\label{tmdung}
\end{eqnarray}
For this case, the gauge link is in the adjoint representation.   

The above definitions are also similar to the non-diffractive TMDs for the semin-inclusive DIS (SIDIS)~\cite{Mulders:1995dh,Boer:1997nt,Bacchetta:2006tn}. However, there is a major difference between these two TMDs. In the non-diffractive TMDs, the associate gauge links can go to either $+\infty$ or $-\infty$~\cite{Collins:2002kn,Ji:2002aa,Belitsky:2002sm}, where the $-\infty$ gauge links correspond to the Drell-Yan type of hard processes, such as virtual photon, $Z/W$ vector Boson, Higgs Boson productions in $pp$ collisions. But, for the diffractive TMDs, the gauge links can only go to $+\infty$, corresponding to the semi-inclusive diffractive DIS processes in $ep$/$eA$ collisions. This is because the QCD factorization is known to break down for the diffractive processes in $pp$ collisions~\cite{DeTar:1974vx,Cardy:1974vq,Collins:1992cv,Collins:2001ga,CDF:2000rua}, where the relevant TMD definition would require a gauge link goes to $-\infty$. In other words, the TMD DPDF with gauge link going to $-\infty$ does not exist.

\begin{figure}[t]
\begin{center}
\includegraphics[width=0.35\textwidth]{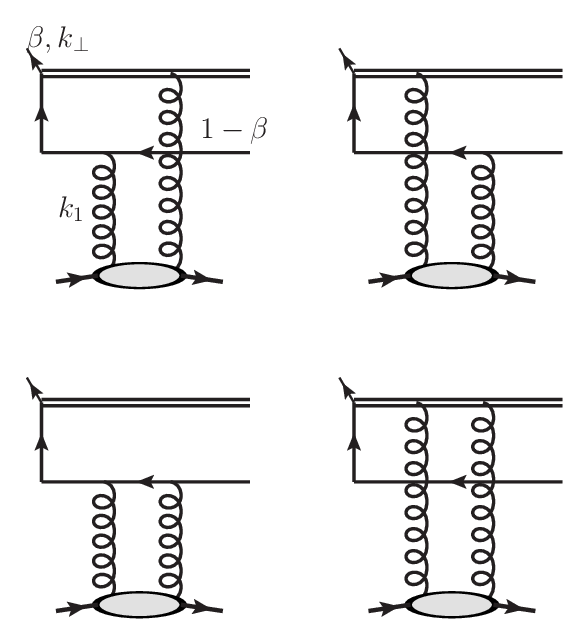}
\end{center}
\caption[*]{Feynman diagrams for the diffractive quark TMDs  at small-$x$ in the dipole formalism.
}
\label{fig:dff}
\end{figure}

At small-$x$, we can compute the TMD DPDFs in the CGC formalism~\cite{Mueller:1993rr,Mueller:1999wm,McLerran:1993ni,McLerran:1993ka,McLerran:1994vd}. The leading order Feynman diagrams have been computed in Ref.~\cite{Hatta:2022lzj}. For convenience, we show the four diagrams for the quark TMD DPDF calculations. Similar diagrams can be drawn for the gluon TMD DPDF. The quark TMD DPDF is related to the color dipole S-matrix 
\begin{eqnarray}
&&\mathcal{F}_x(q_\perp, \Delta_\perp)=\int\frac{d^2b_\perp d^2r_\perp}{(2\pi)^4} e^{iq_\perp \cdot r_\perp +i \Delta_\perp\cdot b_\perp}\nonumber\\
&&\qquad \times \frac{1}{N_{c}}\left\langle\text{Tr}\left[
U\left( b_{\perp }+\frac{r_\perp}{2}\right) U^{\dagger }\left( b_{\perp }-\frac{r_\perp}{2} \right)\right]\right\rangle_x \label{g2} \ , %\\
\end{eqnarray}
where the brackets $\langle ...\rangle$ denote the CGC averaging \cite{McLerran:1993ni} and 
\beq
U(b_\perp)={\rm P}\exp\left(ig\int_{-\infty}^\infty dz^-A^+(z^-,b_\perp)\right), \label{wilson}
\eeq
is the Wilson line along the lightcone in the fundamental representation. Similarly, the gluon TMD DPDF is related to the dipole S-matrix in the adjoint representation. 
It is known that these dipole S-matrices  contain nontrivial correlations between   $\Delta_\perp$ and $q_\perp$, and also the spin vector $\vec{S}_\perp$ if the nucleon is transversely polarized. As already commented in \cite{Hatta:2022lzj}, such correlations will be reflected in TMD DPDFs with observable consequences. 

The goal of this paper is to compute these correlations in the quark and gluon DPDFs in the CGC framework. To do so, we first recapitulate the main result of \cite{Hatta:2022lzj}. The typical Feynman diagrams necessary to evaluate the quark and gluon TMD DPDFs are shown in Fig.~\ref{fig:dff} where the double lines represent gauge links. To the order of accuracy, the final state $|X\rangle$  is saturated by a single antiquark and a gluon  for the quark and gluon DPDFs, respectively. The phase space of the antiquark/gluon is integrated out while ensuring the rapidity gap $Y_{I\!\!P}=\ln 1/x_{I\!\!P}$ due to the  colorless exchange from the nucleon target. 

The result for the quark TMD DPDF is 
\begin{eqnarray}
&&x\frac{d\, f_q^D(\beta,k_\perp;x_{I\!\!P})}{dY_{I\!\!P} dt d\phi_\Delta} =   \int
d^2k_{1\perp}d^2k_{2\perp} {\cal F}_{x_{I\!\!P}}(k_{1\perp},\Delta_\perp)
\nonumber \\ 
&&~~\times{\cal F}_{x_{I\!\!P}}(k_{2\perp},\Delta_\perp)
\frac{N_c\beta}{(2\pi)^2} {\cal T}_q(k_\perp,k_{1\perp},k_{2\perp}) \ ,
\label{dffquark}%\nonumber\\
\end{eqnarray} 
where $\phi_\Delta$ represents the azimuthal angle of the recoil nucleon momentum $\vec{\Delta}_\perp$. $k_{1\perp}$ is the transverse momentum of one of the two vertical gluon lines in the left diagram of Fig.~\ref{fig:dff}, and $k_{2\perp}$ is for the complex conjugate diagram. ${\cal T}_q$ represents the sum of  four terms  ${\cal T}_q\equiv T_q(k_\perp,k_{1\perp},k_{2\perp})-T_q(k_\perp,0,k_{2\perp})-T_q(k_\perp,k_{1\perp},0)+T_q(k_\perp,0,0)$ where 
\begin{eqnarray}
&&    T_q(k_\perp,k_{1\perp},k_{2\perp})=\nonumber\\
&&\frac{ k_{1\perp}'\cdot k_{2\perp}'k_\perp^2}
{\left[\beta k_\perp^2+(1-\beta)k_{1\perp}^{\prime 2}\right]
\left[\beta k_\perp^2+(1-\beta)k_{2\perp}^{\prime 2}\right]}\ , \label{eq:tq}
\end{eqnarray}
with  $k_{i\perp}'=k_\perp-k_{i\perp}$. $\mathcal{F}_{x_{I\!\!P}}(k_{i\perp}, \Delta_\perp)$ is as in (\ref{g2}) evaluated at $x=x_{I\!\!P}$. 
Similarly, the  gluon TMD DPDF reads 
\begin{eqnarray}
&&x\frac{d f_g^D(\beta,k_\perp;x_{I\!\!P})}{dY_{I\!\!P} dt d\phi_\Delta } =  \int
d^2k_{1\perp}d^2k_{2\perp} {\cal G}_{x_{I\!\!P}}(k_{1\perp},\Delta_\perp)
\nonumber \\ 
&&~~\times{\cal G}_{x_{I\!\!P}}(k_{2\perp},\Delta_\perp)\frac{N_c^2-1}{2\pi^2(1-\beta)}  {\cal T}_g(k_\perp,k_{1\perp},k_{2\perp})\ ,\label{eq:dffgluon}
\end{eqnarray}
where we again defined ${\cal T}_g\equiv 
T_g(k_\perp,k_{1\perp},k_{2\perp})-T_g(k_\perp,0,k_{2\perp})-T_g(k_\perp,k_{1\perp},0)+T_g(k_\perp,0,0)$ and  \begin{eqnarray}
&&    T_g(k_\perp,k_{1\perp},k_{2\perp})=\frac{1}{\left[\beta k_\perp^2+(1-\beta)k_{1\perp}^{\prime 2}\right]}\nonumber\\
&&~~~\times \frac{1}{%\left[\beta k_\perp^2+(1-\beta)k_{1\perp}^{\prime 2}\right]
\left[\beta k_\perp^2+(1-\beta)k_{2\perp}^{\prime 2}\right]}\left[\beta(1-\beta){k_\perp^2}\frac{k_{1\perp}^{\prime 2}+k_{2\perp}^{\prime 2}}{2}\right.\nonumber\\
&& ~~~\left.+(1-\beta)^2 (k_{1\perp}'\cdot k_{2\perp}')^2+\beta^2 \frac{(k_\perp^2)^2}{2}\right]
\ . %\label{eq:tg}
\end{eqnarray}  
For a later purpose, we make a one step back in the derivation and write 
\begin{equation}
    T_g(k_\perp,k_{1\perp},k_{2\perp})=\delta_\perp^{\alpha\alpha'}\delta_\perp^{\nu\nu'}R_g^{\alpha \nu}(k_\perp,k_{1\perp})R_g^{\alpha' \nu'}(k_\perp,k_{2\perp})\ , \label{contract}
\end{equation}
    where 
\begin{equation}
    R_g^{\alpha\nu}(k_\perp,k_{i\perp})=\frac{(1-\beta)k_{i\perp}^{\prime \alpha}k_{i\perp}^{\prime \nu}+\beta k_\perp^2\delta_\perp^{\alpha\nu}/2}{(1-\beta)k_{i\perp}^{\prime 2}+\beta k_\perp^2} \ .
\end{equation}
Finally, ${\cal G}$ is the gluon dipole S-matrix 
\begin{eqnarray}
&&\mathcal{G}_x(q_\perp, \Delta_\perp)=\int\frac{d^2b_\perp d^2r_\perp}{(2\pi)^4} e^{iq_\perp \cdot r_\perp +i \Delta_\perp\cdot b_\perp}\nonumber\\
&&\times \frac{1}{N^2_{c}-1}\left\langle\text{Tr}\left[
\widetilde{U}\left( b_{\perp }+\frac{r_\perp}{2}\right) \widetilde{U}^{\dagger }\left( b_{\perp }-\frac{r_\perp}{2} \right)\right]\right\rangle_x \  ,\label{g2g}
\end{eqnarray}
where $\widetilde{U}$ is the same Wilson line but in the adjoint representation. 

\section{Elliptic TMD DPDFs}
Let us first discuss the `elliptic' TMD DPDFs. For this purpose, we can follow the parameterization in Ref.~\cite{Hatta:2016dxp,Pasechnik:2023mdd} 
\begin{eqnarray}
    &&\mathcal{F}_x
    =\mathcal{F}_0(|q_\perp|, |\Delta_\perp|)\!+\! 2\cos(2\phi_q\!-\!2\phi_\Delta)\mathcal{F}_\epsilon(|q_\perp|, |\Delta_\perp|) \,,\nonumber\\ 
      &&\mathcal{G}_x
    =\mathcal{G}_0(|q_\perp|, |\Delta_\perp|)\!+\!2\cos(2\phi_q\!-\! 2\phi_\Delta)\mathcal{G}_\epsilon(|q_\perp|, |\Delta_\perp|). \nn
    \label{f0}
\end{eqnarray}
Therefore, the $\cos(2\phi)$ term in the TMD DPDF can be written as
\begin{eqnarray}
&&x\frac{d\, f_{q\epsilon}^D(\beta,k_\perp;x_{I\!\!P})}{dY_{I\!\!P} dtd\phi_\Delta} =   2\int
d^2k_{1\perp}d^2k_{2\perp} {\cal F}_\epsilon(k_{1\perp},\Delta_\perp)
 \\
&&\times{\cal F}_0(k_{2\perp},\Delta_\perp)
\frac{N_c\beta}{(2\pi)^2} {\cal T}_q(k_\perp,k_{1\perp},k_{2\perp}) 2\cos(2\phi_{k_1}-2\phi_\Delta)\,.
\label{dffquarkepsilon}\nonumber
\end{eqnarray} 
We can perform the angular integral in Eq.~(\ref{dffquark}). For the $\phi_\Delta$-independent term, we have
\begin{eqnarray}
    x\frac{d f_q^D(\beta,k_\perp;x_{I\!\!P})}{dY_{I\!\!P}dt d\phi_\Delta} &=& \frac{ N_c \beta}{16(1-\beta)^2} \left(\Gamma^q(\beta,k_\perp)\right)^2, \\
    x\frac{d f_g^D(\beta,k_\perp;x_{I\!\!P})}{dY_{I\!\!P}dt d\phi_\Delta} &=& \frac{ (N_c^2-1) }{16(1-\beta)^3} \left(\Gamma^g(\beta,k_\perp)\right)^2
    \,,
\end{eqnarray}
where we have taken the limit $\Delta_\perp\to 0$. 
$\Gamma^q(\beta, k_\perp)$ and $\Gamma^g(\beta,k_\perp)$ are defined as
\begin{eqnarray}
    \Gamma^q(\beta,k_\perp)&=&\int_0^\infty dk_{1\perp}^2{\cal F}_0(k_{1\perp}^2) \left\{ 1-2\beta +\frac{\alpha_{1\perp}^2}{\omega_{1\perp}^2}\right\} \label{conv}\\
    \Gamma^g(\beta,k_\perp)&=&\int_0^\infty dk_{1\perp}^2{\cal G}_0(k_{1\perp}^2) \left\{ \frac{2\beta(1-\beta)k_\perp^2}{\omega_{1\perp}^2}-\frac{\omega_{1\perp}^2}{k_\perp^2}\right.\nonumber\\
    &&\left.+\beta^2+(1-\beta)^2+(1-\beta)\frac{k_{1\perp}^2}{k_\perp^2}\right\}\,, \label{conv2}
\end{eqnarray}
where $\alpha_{1\perp}^2=(1-\beta)k_{1\perp}^2-(1-2\beta)k_\perp^2$ and 
\begin{equation}
    \omega_{1\perp}^2=\sqrt{(k_\perp^2+(1-\beta)k_{1\perp}^2)^2-4(1-\beta)^2k_\perp^2 k_{1\perp}^2} \ .
\end{equation}
In the elliptic term, we write ${\cal F}_\epsilon(|q_\perp|,|\Delta_\perp|)\approx \Delta_\perp^2 \tilde{{\cal F}}_\epsilon(q_\perp^2)$ as $\Delta_\perp\to 0$ and perform the angular integral  
\beq
    x\frac{d f_{q\epsilon}^D(\beta,k_\perp;x_{I\!\!P})}{dY_{I\!\!P}dt d\phi_\Delta}\!\! \!&=&\! \!\!\frac{ N_c \beta \Delta_\perp^2}{16(1-\beta)^2} \Gamma^q\Gamma_\epsilon^q\cos (2\phi_k-2\phi_\Delta)\,,\\
       x\frac{d f_{g\epsilon}^D(\beta,k_\perp;x_{I\!\!P})}{dY_{I\!\!P}dt d\phi_\Delta}\!\! \!&=&\!\!\! \frac{ (N_c^2-1) \Delta_\perp^2}{16(1-\beta)^3} \Gamma^g\Gamma_\epsilon^g\cos (2\phi_k-2\phi_\Delta) \,, \nn
    \eeq
where $\Gamma^{q,g}(\beta,k_\perp)$ are the same as above and
\begin{eqnarray}
 &&   \Gamma_\epsilon^q(\beta,k_\perp)=2\int_0^\infty dk_{1\perp}^2\tilde{{\cal F}}_\epsilon(k_{1\perp}^2)\frac{\alpha_{1\perp}^2}{k_\perp^2k_{1\perp}^2(1-\beta)^2}\nonumber\\
    &&\times \left\{\omega_{1\perp}^2-k_\perp^2-(1-\beta)k_{1\perp}^2+\frac{2(1-\beta)^2k_\perp^2k_{1\perp}^2}{\omega_{1\perp}^2}\right\}\,, \label{el1}\\
 &&   \Gamma_\epsilon^g(\beta,k_\perp)=2\int_0^\infty dk_{1\perp}^2\tilde{{\cal G}}_\epsilon(k_{1\perp}^2)\frac{\omega_{1\perp}^4-2\beta(1-\beta)k_\perp^4}{k_\perp^4k_{1\perp}^2(1-\beta)^2}\nonumber\\
    &&~~\times \left\{\frac{2(1-\beta)^2k_\perp^2k_{1\perp}^2}{\omega_{1\perp}^2}+\left(k_\perp^2+(1-\beta)k_{1\perp}^2\right)\nonumber\right.\\
    &&~~\times \left.\left(1-\frac{(k_\perp^2+(1-\beta)k_{1\perp}^2}{\omega_{1\perp}^2}\right)\right\}\ . \label{el2}
\end{eqnarray}

\begin{figure}[t]
\begin{center}
\includegraphics[width=0.35\textwidth]{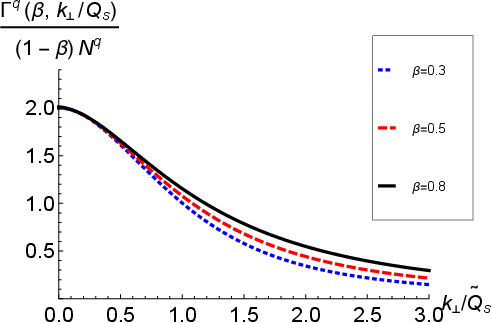}
\vskip 1cm
\includegraphics[width=0.35\textwidth]{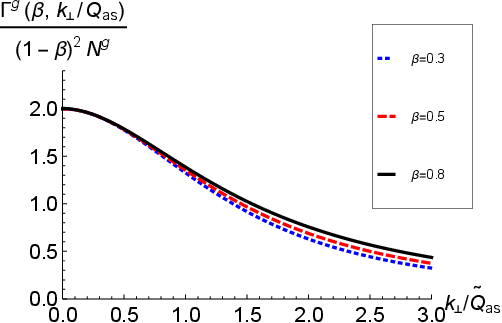}
\end{center}
\caption[*]{Azimuthal angular symmetric contributions to $\Gamma^{q,g}$ the TMD DPDFs as functions of $k_\perp/\widetilde Q_{(a)s}$ where $\widetilde Q_{(a)s}=\sqrt{1-\beta}Q_{(a)s}$. 
}
\label{fig:gammaqg}
\end{figure}

At small-$x$, it is known that ${\cal F}_0$ and ${\cal G}_0$ show the geometric scaling behavior. Namely, they depend on $x$ and $k_\perp$ through the combination $k_\perp/Q_{(a)s}(x)$ where $Q_{s}$ and $Q_{as}$ with $Q_{as}^2=\frac{N_c}{C_F}Q_{s}^2$ are the quark and gluon saturation scales, respectively. Such a scaling is inherited by the ordinally quark and gluon TMDs at small-$x$.  
However, this is not the case for TMD DPDFs  because of the additional parameter $\beta$. Nevertheless, it has been observed in \cite{Iancu:2021rup,Hatta:2022lzj} that TMD DPDFs exhibit the modified geometric scaling in terms of the rescaled variable  $k_\perp/\widetilde{Q}_{(a)s}$ where $\widetilde{Q}_{(a)s}\equiv \sqrt{1-\beta} Q_{(a)s}$. 

Previous studies  \cite{Hagiwara:2016kam,Yao:2018vcg} have shown that geometric scaling is violated in the elliptic, or more generally, angular-dependent gluon   distributions due to the small-$x$ evolution effect. However, in regimes where the evolution is not important, we may expect that there is an approximate (modified) geometric scaling. We study this by using a simple Gaussian model  
\beq 
{\cal F}_0(q_\perp),\ {\cal G}_0(q_\perp) &=& \frac{S_\perp}{(2\pi)^2} \frac{1}{\pi Q_{(a)s}^2}  e^{-q_\perp^2/Q_{(a)s}^2}\,,   \label{fepsilon}\\
{\cal F}_\epsilon(|q_\perp|,|\Delta_\perp|)&=&\frac{S_\perp}{(2\pi)^2} \frac{\Delta_\perp^2q_\perp^2 }{\pi(Q_{s}^2)^2\Lambda^2 } e^{-q_\perp^2/Q_{s}^2}\,, \label{feps}
\\
{\cal  G}_\epsilon(|q_\perp|,|\Delta_\perp|)&=&\frac{S_\perp}{(2\pi)^2}  \frac{\Delta_\perp^2q_\perp^2 }{\pi(Q_{as}^2)^2\Lambda^2}  e^{-q_\perp^2/Q_{as}^2}\,,   \label{geps}
\eeq 
 where $S_\perp$ is the transverse area of the nucleon. 
In the elliptic part, we have implemented the minimum requirement that ${\cal F}_\epsilon,{\cal G}_\epsilon \propto q_\perp^2\Delta_\perp^2$ as $q_\perp,\Delta_\perp\to 0$. 
For more realistic models, see  \cite{Hagiwara:2016kam,Zhou:2016rnt,Iancu:2017fzn,Linek:2023kga,Pasechnik:2023mdd}. In $r_\perp$-space, (\ref{fepsilon}) and (\ref{feps}) implies 
\beq
&&\left\langle \frac{1}{N_c}{\rm Tr}U\left(\frac{r_\perp}{2}\right)U^\dagger\left(-\frac{r_\perp}{2}\right)\right\rangle \nn
&& = e^{-\frac{r_\perp^2 Q_s^2}{4}}\left(1-2\cos (2\phi_r-2\phi_\Delta)\frac{Q_s^2r_\perp^2\Delta_\perp^2}{4\Lambda^2}\right).
\eeq
(\ref{geps}) is then consistent with the identity 
 ${\rm Tr}[\tilde{U}\tilde{U}^\dagger]=|{\rm Tr} [UU^\dagger]|^2-1$ and the relation $Q_{as}^2\approx 2Q_s^2$ in the large-$N_c$ approximation.

\begin{figure}[t]
\begin{center}
\includegraphics[width=0.35\textwidth]{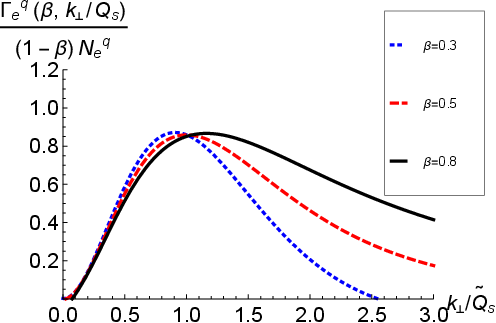}
\vskip 1cm
\includegraphics[width=0.35\textwidth]{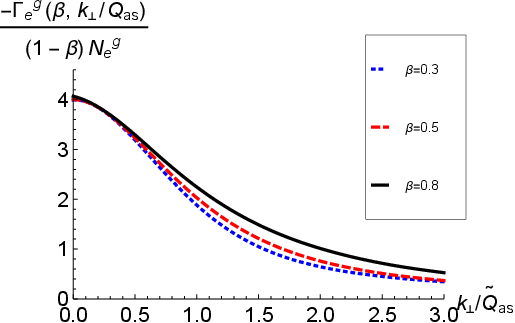}
\end{center}
\caption[*]{Azimuthal angular asymmetric contribution to the elliptic TMD DPDFs $\Gamma_\epsilon^{q,g}$ as functions of $k_\perp/\widetilde{Q}_{(a)s}$.
}
\label{fig:gammaeqg}
\end{figure}

With these parametrizations, the integrals (\ref{conv}), (\ref{conv2}), (\ref{el1}), (\ref{el2}) can be readily performed. We first extract the analytical behavior as  $\beta\to 1$. As explained in \cite{Hatta:2022lzj}, this can be conveniently done by rescaling $k_{1\perp}^2=\tilde{k}^2_{1\perp}/(1-\beta)$ and expanding around $\beta=1$. The result is 
\beq
\Gamma^q,\Gamma^q_\epsilon, \Gamma^g_\epsilon  \propto 1-\beta, \qquad 
\Gamma^g \propto (1-\beta)^2\,.
\eeq
We then plot the rescaled distributions  $\Gamma^q/(1-\beta)N^q$ with normalization factors $N^q=N^g=S_\perp/4\pi^3$ and $N_q^\epsilon=N_g^\epsilon=S_\perp/4\pi^3\Lambda^2$, etc.  as functions of $k_\perp/\widetilde{Q}_{(a)s}$ for three different values $\beta=$0.3, 0.5, 0.8.   We see in Fig.~\ref{fig:gammaqg} that the azimuthally symmetric distributions approximately obey the modified geometric scaling consistently with the previous result \cite{Hatta:2022lzj}.  
On the other hand, we see in Fig.~\ref{fig:gammaeqg} that   the elliptic quark distribution $\Gamma^q_\epsilon$ does not show the scaling behavior at high $k_\perp$. The elliptic gluon distribution $\Gamma^g_\epsilon$ approximately does, but somewhat unexpectedly, it does not go to zero as $k_\perp \to 0$. Normally one would expect that the $\cos 2\phi_k$ angular dependence should vanish as $k_\perp^2$ as $k_\perp\to 0$. While this does not cause immediate problems, it is still an intriguing observation.

\section{Spin-dependent TMD DPDF}

In order to study spin-dependent TMD DPDFs, we introduce the nucleon helicity ($\pm$) dependent dipole amplitudes. ${\cal F}_0$ in (\ref{f0}) can be interpreted as the helicity-conserving  amplitude 
\begin{eqnarray}
    &&\mathcal{F}_x^{++}(q_\perp,\Delta_\perp)=\mathcal{F}_x^{--}(q_\perp,\Delta_\perp)\nn 
    &&
=\mathcal{F}_0(|q_\perp|,|\Delta_\perp|)+i\Delta_\perp\cdot q_\perp O(|q_\perp|,|\Delta_\perp|)+\cdots\,. \label{indep}
\end{eqnarray}
The imaginary piece is sometimes referred to as the spin-independent odderon, see, e.g., \cite{Kovchegov:2003dm,Hatta:2005as,Dumitru:2018vpr,Benic:2023ybl}.  
On the other hand, the helicity-flip dipole amplitude can be parameterized as 
\begin{eqnarray}
    \mathcal{F}_x^{+-}(q_\perp, \Delta_\perp)&=& \frac{\Delta_\perp^x+i\Delta_\perp^y}{M} {\cal F}_{1T} \label{dep} \\
    && -i\frac{q_\perp^x+iq_\perp^y}{M}O_{1T}^\perp(|q_\perp|,|\Delta_\perp|)+\cdots \,, \nonumber 
\end{eqnarray}
where $M$ is the nucleon mass. 
$O_{1T}^\perp$ is the so-called spin-dependent odderon~\cite{Zhou:2013gsa, Boer:2015pni}. 
 We have adopted the following conventions for the transverse polarization states
\begin{eqnarray}
    |\uparrow \downarrow\rangle_x &=& \frac{1}{\sqrt{2}}\left(|+\rangle\pm |-\rangle\right)\,,\\
    |\uparrow \downarrow\rangle_y &=& \frac{1}{\sqrt{2}}\left(|+\rangle\pm i|-\rangle\right)\,,
\end{eqnarray}
where the left hand sides represent transversely polarized nucleon states along the $\hat x$ and $\hat y$ directions, respectively. 
The spin-dependent odderon is known to contribute to the Sivers function for the non-diffractive SIDIS, leading to non-zero gluon Sivers function as small-$x$. Meanwhile, it also contributes to the spin asymmetries in the exclusive processes~\cite{Boussarie:2019vmk}. Substituting the above expression to the quark TMD DPDF, we will show that it will contribute to the Sivers asymmetry in the diffractive semi-inclusive processes.

We follow the same parameterization as that of non-diffractive TMDs for the transverse spin dependent TMD DPDF,
\begin{equation}
\delta f^D_q(\vec{S}_\perp)=\frac{k_\perp \sin(\phi_k-\phi_S)}{M} f_{1Tq}^{D\perp}(\beta,k_\perp;x_{I\!\!P}) \ ,
\end{equation}
where $\phi_S$ is the angle of the transverse nucleon vector $\vec{S}_\perp$. 
To zeroth order in $\Delta_\perp$, the quark Sivers TMD DPDF can be evaluated as
\begin{eqnarray}
&&x\frac{d\, f_{1Tq}^{D\perp}(\beta,k_\perp;x_{I\!\!P})}{dY_{I\!\!P} dt  } \sin(\phi_k-\phi_S) \nonumber\\
&&~~~~~~~~=   2\int 
d^2k_{1\perp}d^2k_{2\perp} O_{1T}^\perp(k_{1\perp},\Delta_\perp)\nonumber
 \\
&&~~~~~~~~~~~\times{\cal F}_0(k_{2\perp},\Delta_\perp)
\frac{N_c\beta}{2\pi} {\cal T}_q(k_\perp,k_{1\perp},k_{2\perp})\nonumber\\
&&~~~~~~~~~~~\times \frac{k_{1\perp}}{k_\perp}\sin(\phi_{k_1}-\phi_S) \,,
\label{dffquarksivers}
\end{eqnarray} 
where we summed over the final state nucleon spin $S'$ and integrated over $\phi_\Delta$. 
Again, after integrating over $\phi_{k_1}$ and $\phi_{k_2}$  we find a Sivers-type distribution  
\begin{eqnarray}
    x\frac{d f_{1Tq}^{D\perp}(\beta,k_\perp;x_{I\!\!P})}{dY_{I\!\!P}dt } = \frac{\pi N_c \beta}{8(1-\beta)^2} \Gamma^q\Gamma_{S_\perp}^q\ , 
\end{eqnarray}
where $\Gamma^{q}(k_\perp,\beta)$ is the same as above and 
\begin{eqnarray}
\Gamma_{S_\perp}^q(k_\perp,\beta)&=&\frac{-1}{1-\beta}\int_0^\infty dk_{1\perp}^2O_{1T}^\perp(k_{1\perp}^2)\frac{\alpha_{1\perp}^2}{k_\perp^2}\nonumber\\
    &&\times \left\{1-\frac{k_\perp^2+(1-\beta)k_{1\perp}^2}{\omega_{1\perp}^2}\right\}\,,
\end{eqnarray}
with $\Gamma_{S_\perp}^q\sim (1-\beta)^0$ as $\beta \to 1$.

\begin{figure}[t]
\begin{center}
\includegraphics[width=0.35\textwidth]{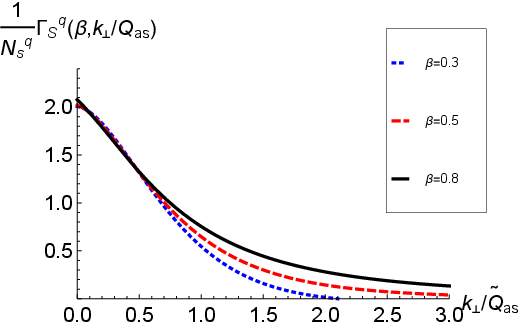}
\end{center}
\caption[*]{Sivers contribution to the TMD quark DPDFs $\Gamma_S^{q}$ as functions of $k_\perp/\widetilde{Q}_s$.
}
\label{fig:gammasq}
\end{figure}

To illustrate the above feature, we assume a saturation model for $O_{1T}^\perp(q_\perp)$,
\begin{equation}
    O_{1T}^\perp(q_\perp;\Delta_\perp=0)=\frac{S_\perp}{(2\pi)^2}\frac{c_{odd}}{\pi Q_s^2}e^{-{q_\perp^2}/{Q_s^2}}\ ,
\end{equation}
where we introduced a parameter $c_{odd}$ to represent the small contribution from the Odderon.  In Fig.~\ref{fig:gammasq}, we show $\Gamma_{S_\perp}^q(\beta,k_\perp)/N_S^q$ with $N_S^q=c_{odd}S_\perp/(2\pi)^2$ as function of $k_\perp/\widetilde{Q}_s$ for three different $\beta$ values. 
Again, the quark Sivers DPDF contribution $\Gamma_{S_\perp}^q(\beta,k_\perp)$ has a modified geometric scaling behavior below the saturation scale $\widetilde{Q}_s$. However, at large transverse momentum, the scaling behavior is broken.

At higher orders in $\Delta_\perp$, more complicated angular correlations which involve all three angles $\phi_{k,S,\Delta}$ are possible. For example,  the ${\cal O}(\Delta_\perp)$ terms in (\ref{indep}) and (\ref{dep}) lead to  
\beq
f_q^D\sim(\vec{S}_\perp \!\times \! \vec{\Delta}_\perp)( \vec{\Delta}_\perp \cdot \vec{k}_\perp) 
\sim \Delta_\perp^2 \sin(\phi_S+\phi_k-2\phi_\Delta)\,.
\eeq

Interestingly, the above spin-dependent correlations are entirely absent in the gluon TMD DPDF. This is because the dipole S-matrix in the adjoint representation (\ref{g2g}) is a real operator whereas the one in the fundamental representation has a nonzero imaginary part which can be identified as the odderon. 
In contrast, in the TMD case  both the quark and gluon Sivers functions are related to the spin-dependent odderon at small-$x$.

\section{Vanishing of the Linearly Polarized Gluon TMD DPDF}

Next we consider the generalization of the linearly polarized gluon TMD to DPDF. This is defined by replacing in (\ref{tmdung})
\beq
 \langle F^{+\mu}...F_\mu^{\ +}\rangle &\to& \langle F^{+i}...F^{+j}\rangle \nn
&=& \frac{\delta^{ij}}{2}f_g^D + \left(\frac{k_\perp^ik_\perp^j}{k_\perp^2}-\frac{\delta^{ij}}{2}\right) h_{1g}^{\perp D}.
\eeq
It is straightforward to compute the linearly polarized gluon TMD DPDF $h_{1g}^{\perp D}$. All we need to do is to replace (\ref{contract}) with 
\begin{equation}
T_g^h(k_\perp,k_{1\perp},k_{2\perp})=\delta_\perp^{\alpha\alpha'}\delta_{h\perp}^{\nu\nu'}R_g^{\alpha \nu}(k_\perp,k_{1\perp})R_g^{\alpha' \nu'}(k_\perp,k_{2\perp})\ ,
\end{equation}
where
$\delta_{h\perp}^{\nu\nu'}=\frac{2k_\perp^\nu k_\perp^{\nu'}}{k_\perp^2}-\delta_\perp^{\nu\nu'}$. This gives 
\begin{eqnarray}
&&x\frac{d h_{1g}^{\perp D}(\beta,k_\perp;x_{I\!\!P})}{dY_{I\!\!P} dt } =  \int
d^2k_{1\perp}d^2k_{2\perp} {\cal G}_{x_{I\!\!P}}(k_{1\perp},\Delta_\perp)
\nonumber \\ 
&&~~\times{\cal G}_{x_{I\!\!P}}(k_{2\perp},\Delta_\perp)\frac{N_c^2-1}{\pi(1-\beta)}  {\cal T}_g^h(k_\perp,k_{1\perp},k_{2\perp})\ ,%\label{eq:dffgluon}
\end{eqnarray}
where we again defined ${\cal T}_g^h\equiv 
T_g^h(k_\perp,k_{1\perp},k_{2\perp})-T_g^h(k_\perp,0,k_{2\perp})-T_g^h(k_\perp,k_{1\perp},0)+T_g^h(k_\perp,0,0)$. Explicitly,  
\begin{eqnarray}
&&    T_g^h(k_\perp,k_{1\perp},k_{2\perp})=-\frac{1-\beta}{2k_\perp^2}\frac{1}{\left[\beta k_\perp^2+(1-\beta)k_{1\perp}^{\prime 2}\right]}\nonumber\\
&&~~~\times \frac{1}{
\left[\beta k_\perp^2+(1-\beta)k_{2\perp}^{\prime 2}\right]}\left[\beta{(k_\perp^2)^2}\left({k_{1\perp}^{\prime 2}+k_{2\perp}^{\prime 2}}\right)\right.\nonumber\\
&& ~~~-4(1-\beta)k_\perp\cdot k_{1\perp}' k_\perp\cdot k_{2\perp}' k_{1\perp}'\cdot k_{2\perp}'\nonumber\\
&& ~~~-2\beta k_\perp^2((k_\perp\cdot k_{1\perp}')^2+(k_\perp\cdot k_{2\perp}')^2)\nonumber\\
&& ~~~\left.
+2(1-\beta) k_\perp^2(k_{1\perp}'\cdot k_{2\perp}')^2
\right]
\ . \label{eq:tg}
\end{eqnarray}
Interestingly, however, we find that ${\cal T}_g^h$ identically vanishes due to a highly nontrivial cancellation among the four terms in ${\cal T}_g^h$. This indicates that the two-gluon exchange ``Pomeron" does not couple to the linearly polarized gluon distribution and unpolarized gluon in the final state. This is different from the non-diffractive TMD gluon distribution, which comes from quark radiation in the CGC and contributes to a significant linearly polarization for both dipole and Weizsaecker-Williams gluon distributions~\cite{Metz:2011wb}.

\section{Conclusion}

In this paper, we have computed various transverse momentum dependent diffractive PDFs in the small-$x$ dipole formalism. We focused on, in particular, the nontrivial correlations between the transverse momentum $\vec{k}_\perp$ of the probing quark/gluon, the momentum transfer $\vec{\Delta}_\perp$ from the nucleon target, and the transverse polarization vector $\vec{S}_\perp$. Interesting results were found from our derivations. First, the linearly polarized gluon TMD does not exist in the diffractive case. Second, the Sivers and elliptic TMD DPDFs have the `modified geometric scaling' behavior \cite{Iancu:2021rup,Hatta:2022lzj}. However, the scaling of these angular-dependent DPDFs is not as clear as in  their angular-independent counterparts, and in some cases explicitly violated in the large-$k_\perp$ region.  We also expect that the remaining scaling property is washed out  once the small-$x$ evolution effects are  turned on \cite{Hagiwara:2016kam,Yao:2018vcg}. 
Finally, the angular-dependent DPDFs have different large-$\beta$ behaviors (different powers of $1-\beta$) from the angular independent distributions. 

It will be interesting to investigate the phenomenological implications of our findings for future measurements at the EIC such as semi-inclusive hadron/jet production in diffractive DIS  (Fig.~\ref{fig:siddis}) and two plus one jet diffractive production. 
In addition, it will be instructive to study inclusive or semi-inclusive diffractive processes in the collinear generalized parton distribution (GPD) framework.  Some preliminary exploration has been carried out in Ref.~\cite{Guo:2023uis}. More systematic studies are needed to fully explore the potential of this physics at the EIC. We will come back to this in a future publication. \\

{\bf Acknowledgments:} We thank Shohini Bhattacharya and Edmond Iancu for discussions. Y.~H. thanks Lawrence Berkeley National Laboratory, where this work was initiated, for hospitality. 
F.~Y. is grateful to the support from France-Berkeley-Fund from University of California at Berkeley. This material is based upon work supported by the LDRD programs of  Brookhaven Science Associates, and by the U.S. Department of Energy, Office of Science, Office of Nuclear Physics, under contract numbers DE-AC02-05CH11231 and DE-SC0012704. It is also  under the umbrella of the Quark-Gluon Tomography (QGT) Topical Collaboration with Award DE-SC0023646 and  the Saturated Glue (SURGE) Topical Theory Collaboration.


\begin{thebibliography}{99}
%\cite{Bartels:1996ne}
\bibitem{Bartels:1996ne}
J.~Bartels, H.~Lotter and M.~W\"usthoff,
%``Quark-antiquark production in DIS diffractive dissociation,''
Phys. Lett. B \textbf{379}, 239-248 (1996)
[erratum: Phys. Lett. B \textbf{382}, 449-449 (1996)]
doi:10.1016/0370-2693(96)00412-1
[arXiv:hep-ph/9602363 [hep-ph]].
%106 citations counted in INSPIRE as of 29 Feb 2024

%\cite{Wusthoff:1997fz}
\bibitem{Wusthoff:1997fz}
M.~Wusthoff,
%``Large rapidity gap events in deep inelastic scattering,''
Phys. Rev. D \textbf{56}, 4311-4321 (1997)
doi:10.1103/PhysRevD.56.4311
[arXiv:hep-ph/9702201 [hep-ph]].
%107 citations counted in INSPIRE as of 19 Mar 2024

%\cite{Buchmuller:1998jv}
\bibitem{Buchmuller:1998jv}
W.~Buchmuller, T.~Gehrmann and A.~Hebecker,
%``Inclusive and diffractive structure functions at small x,''
Nucl. Phys. B \textbf{537}, 477-500 (1999)
doi:10.1016/S0550-3213(98)00682-8
[arXiv:hep-ph/9808454 [hep-ph]].
%97 citations counted in INSPIRE as of 29 Feb 2024

%\cite{Golec-Biernat:1999qor}
\bibitem{Golec-Biernat:1999qor}
K.~J.~Golec-Biernat and M.~Wusthoff,
%``Saturation in diffractive deep inelastic scattering,''
Phys. Rev. D \textbf{60}, 114023 (1999)
doi:10.1103/PhysRevD.60.114023
[arXiv:hep-ph/9903358 [hep-ph]].
%967 citations counted in INSPIRE as of 19 Mar 2024

%\cite{Kovchegov:1999ji}
\bibitem{Kovchegov:1999ji}
Y.~V.~Kovchegov and E.~Levin,
%``Diffractive dissociation including multiple pomeron exchanges in high parton density QCD,''
Nucl. Phys. B \textbf{577}, 221-239 (2000)
doi:10.1016/S0550-3213(00)00125-5
[arXiv:hep-ph/9911523 [hep-ph]].
%115 citations counted in INSPIRE as of 18 Mar 2024

%\cite{Kowalski:2006hc}
\bibitem{Kowalski:2006hc}
H.~Kowalski, L.~Motyka and G.~Watt,
%``Exclusive diffractive processes at HERA within the dipole picture,''
Phys. Rev. D \textbf{74}, 074016 (2006)
doi:10.1103/PhysRevD.74.074016
[arXiv:hep-ph/0606272 [hep-ph]].
%552 citations counted in INSPIRE as of 21 Mar 2024

%\cite{Hatta:2006hs}
\bibitem{Hatta:2006hs}
Y.~Hatta, E.~Iancu, C.~Marquet, G.~Soyez and D.~N.~Triantafyllopoulos,
%``Diffusive scaling and the high-energy limit of deep inelastic scattering in QCD at large N(c),''
Nucl. Phys. A \textbf{773}, 95-155 (2006)
doi:10.1016/j.nuclphysa.2006.04.003
[arXiv:hep-ph/0601150 [hep-ph]].
%103 citations counted in INSPIRE as of 29 Feb 2024

%\cite{Marquet:2007nf}
\bibitem{Marquet:2007nf}
C.~Marquet,
%``A Unified description of diffractive deep inelastic scattering with saturation,''
Phys. Rev. D \textbf{76}, 094017 (2007)
doi:10.1103/PhysRevD.76.094017
[arXiv:0706.2682 [hep-ph]].
%99 citations counted in INSPIRE as of 20 Mar 2024

%\cite{Kowalski:2008sa}
\bibitem{Kowalski:2008sa}
H.~Kowalski, T.~Lappi, C.~Marquet and R.~Venugopalan,
%``Nuclear enhancement and suppression of diffractive structure functions at high energies,''
Phys. Rev. C \textbf{78}, 045201 (2008)
doi:10.1103/PhysRevC.78.045201
[arXiv:0805.4071 [hep-ph]].
%126 citations counted in INSPIRE as of 19 Mar 2024

%\cite{Altinoluk:2015dpi}
\bibitem{Altinoluk:2015dpi}
T.~Altinoluk, N.~Armesto, G.~Beuf and A.~H.~Rezaeian,
%``Diffractive Dijet Production in Deep Inelastic Scattering and Photon-Hadron Collisions in the Color Glass Condensate,''
Phys. Lett. B \textbf{758}, 373-383 (2016)
doi:10.1016/j.physletb.2016.05.032
[arXiv:1511.07452 [hep-ph]].
%85 citations counted in INSPIRE as of 25 Mar 2024

%\cite{Hatta:2016dxp}
\bibitem{Hatta:2016dxp}
Y.~Hatta, B.~W.~Xiao and F.~Yuan,
%``Probing the Small- x Gluon Tomography in Correlated Hard Diffractive Dijet Production in Deep Inelastic Scattering,''
Phys. Rev. Lett. \textbf{116}, no.20, 202301 (2016)
doi:10.1103/PhysRevLett.116.202301
[arXiv:1601.01585 [hep-ph]].
%156 citations counted in INSPIRE as of 29 Feb 2024

%\cite{Boussarie:2016ogo}
\bibitem{Boussarie:2016ogo}
R.~Boussarie, A.~V.~Grabovsky, L.~Szymanowski and S.~Wallon,
%``On the one loop $ {\gamma}^{\left(\ast \right)}\to q\overline{q} $ impact factor and the exclusive diffractive cross sections for the production of two or three jets,''
JHEP \textbf{11}, 149 (2016)
doi:10.1007/JHEP11(2016)149
[arXiv:1606.00419 [hep-ph]].
%87 citations counted in INSPIRE as of 20 Mar 2024

%\cite{Boussarie:2019ero}
\bibitem{Boussarie:2019ero}
R.~Boussarie, A.~V.~Grabovsky, L.~Szymanowski and S.~Wallon,
%``Towards a complete next-to-logarithmic description of forward exclusive diffractive dijet electroproduction at HERA: real corrections,''
Phys. Rev. D \textbf{100}, no.7, 074020 (2019)
doi:10.1103/PhysRevD.100.074020
[arXiv:1905.07371 [hep-ph]].
%35 citations counted in INSPIRE as of 25 Mar 2024

%\cite{Fucilla:2022wcg}
\bibitem{Fucilla:2022wcg}
M.~Fucilla, A.~V.~Grabovsky, E.~Li, L.~Szymanowski and S.~Wallon,
%``NLO computation of diffractive di-hadron production in a saturation framework,''
JHEP \textbf{03}, 159 (2023)
doi:10.1007/JHEP03(2023)159
[arXiv:2211.05774 [hep-ph]].
%22 citations counted in INSPIRE as of 26 Mar 2024

%\cite{Zhou:2016rnt}
\bibitem{Zhou:2016rnt}
J.~Zhou,
%``Elliptic gluon generalized transverse-momentum-dependent distribution inside a large nucleus,''
Phys. Rev. D \textbf{94}, no.11, 114017 (2016)
doi:10.1103/PhysRevD.94.114017
[arXiv:1611.02397 [hep-ph]].
%56 citations counted in INSPIRE as of 07 Mar 2024

%\cite{Hagiwara:2017fye}
\bibitem{Hagiwara:2017fye}
Y.~Hagiwara, Y.~Hatta, R.~Pasechnik, M.~Tasevsky and O.~Teryaev,
%``Accessing the gluon Wigner distribution in ultraperipheral $pA$ collisions,''
Phys. Rev. D \textbf{96}, no.3, 034009 (2017)
doi:10.1103/PhysRevD.96.034009
[arXiv:1706.01765 [hep-ph]].
%77 citations counted in INSPIRE as of 25 Mar 2024

%\cite{Mantysaari:2019csc}
\bibitem{Mantysaari:2019csc}
H.~M\"antysaari, N.~Mueller and B.~Schenke,
%``Diffractive Dijet Production and Wigner Distributions from the Color Glass Condensate,''
Phys. Rev. D \textbf{99}, no.7, 074004 (2019)
doi:10.1103/PhysRevD.99.074004
[arXiv:1902.05087 [hep-ph]].
%73 citations counted in INSPIRE as of 20 Mar 2024

%\cite{Mantysaari:2019hkq}
\bibitem{Mantysaari:2019hkq}
H.~M\"antysaari, N.~Mueller, F.~Salazar and B.~Schenke,
%``Multigluon Correlations and Evidence of Saturation from Dijet Measurements at an Electron-Ion Collider,''
Phys. Rev. Lett. \textbf{124}, no.11, 112301 (2020)
doi:10.1103/PhysRevLett.124.112301
[arXiv:1912.05586 [nucl-th]].
%58 citations counted in INSPIRE as of 20 Mar 2024

%\cite{Iancu:2021rup}
\bibitem{Iancu:2021rup}
E.~Iancu, A.~H.~Mueller and D.~N.~Triantafyllopoulos,
%``Probing Parton Saturation and the Gluon Dipole via Diffractive Jet Production at the Electron-Ion Collider,''
Phys. Rev. Lett. \textbf{128}, no.20, 202001 (2022)
doi:10.1103/PhysRevLett.128.202001
[arXiv:2112.06353 [hep-ph]].
%34 citations counted in INSPIRE as of 26 Mar 2024

%\cite{Iancu:2022lcw}
\bibitem{Iancu:2022lcw}
E.~Iancu, A.~H.~Mueller, D.~N.~Triantafyllopoulos and S.~Y.~Wei,
%``Gluon dipole factorisation for diffractive dijets,''
JHEP \textbf{10}, 103 (2022)
doi:10.1007/JHEP10(2022)103
[arXiv:2207.06268 [hep-ph]].
%18 citations counted in INSPIRE as of 20 Mar 2024

%\cite{Hauksson:2024bvv}
\bibitem{Hauksson:2024bvv}
S.~Hauksson, E.~Iancu, A.~H.~Mueller, D.~N.~Triantafyllopoulos and S.~Y.~Wei,
%``TMD factorisation for diffractive jets in photon-nucleus interactions,''
[arXiv:2402.14748 [hep-ph]].
%1 citations counted in INSPIRE as of 20 Mar 2024

%\cite{Hatta:2022lzj}
\bibitem{Hatta:2022lzj}
Y.~Hatta, B.~W.~Xiao and F.~Yuan,
%``Semi-inclusive diffractive deep inelastic scattering at small x,''
Phys. Rev. D \textbf{106}, no.9, 094015 (2022)
doi:10.1103/PhysRevD.106.094015
[arXiv:2205.08060 [hep-ph]].
%24 citations counted in INSPIRE as of 19 Mar 2024

%\cite{Berera:1995fj}
\bibitem{Berera:1995fj}
A.~Berera and D.~E.~Soper,
%``Behavior of diffractive parton distribution functions,''
Phys. Rev. D \textbf{53}, 6162-6179 (1996)
doi:10.1103/PhysRevD.53.6162
[arXiv:hep-ph/9509239 [hep-ph]].
%136 citations counted in INSPIRE as of 19 Mar 2024

%\cite{Mueller:1993rr}
\bibitem{Mueller:1993rr}
A.~H.~Mueller,
%``Soft gluons in the infinite momentum wave function and the BFKL pomeron,''
Nucl. Phys. B \textbf{415}, 373-385 (1994)
doi:10.1016/0550-3213(94)90116-3
%1075 citations counted in INSPIRE as of 18 Mar 2024

%\cite{Mueller:1999wm}
\bibitem{Mueller:1999wm}
A.~H.~Mueller,
%``Parton saturation at small x and in large nuclei,''
Nucl. Phys. B \textbf{558}, 285-303 (1999)
doi:10.1016/S0550-3213(99)00394-6
[arXiv:hep-ph/9904404 [hep-ph]].
%354 citations counted in INSPIRE as of 29 Feb 2024

%\cite{McLerran:1993ni}
\bibitem{McLerran:1993ni}
L.~D.~McLerran and R.~Venugopalan,
%``Computing quark and gluon distribution functions for very large nuclei,''
Phys. Rev. D \textbf{49}, 2233-2241 (1994)
doi:10.1103/PhysRevD.49.2233
[arXiv:hep-ph/9309289 [hep-ph]].
%2310 citations counted in INSPIRE as of 20 Mar 2024

%\cite{McLerran:1993ka}
\bibitem{McLerran:1993ka}
L.~D.~McLerran and R.~Venugopalan,
%``Gluon distribution functions for very large nuclei at small transverse momentum,''
Phys. Rev. D \textbf{49}, 3352-3355 (1994)
doi:10.1103/PhysRevD.49.3352
[arXiv:hep-ph/9311205 [hep-ph]].
%1722 citations counted in INSPIRE as of 25 Mar 2024

%\cite{McLerran:1994vd}
\bibitem{McLerran:1994vd}
L.~D.~McLerran and R.~Venugopalan,
%``Green's functions in the color field of a large nucleus,''
Phys. Rev. D \textbf{50}, 2225-2233 (1994)
doi:10.1103/PhysRevD.50.2225
[arXiv:hep-ph/9402335 [hep-ph]].
%1155 citations counted in INSPIRE as of 18 Mar 2024

%\cite{Hautmann:1998xn}
\bibitem{Hautmann:1998xn}
F.~Hautmann, Z.~Kunszt and D.~E.~Soper,
%``Diffractive deeply inelastic scattering of hadronic states with small transverse size,''
Phys. Rev. Lett. \textbf{81}, 3333-3336 (1998)
doi:10.1103/PhysRevLett.81.3333
[arXiv:hep-ph/9806298 [hep-ph]].
%62 citations counted in INSPIRE as of 29 Feb 2024

%\cite{Hautmann:2000pw}
\bibitem{Hautmann:2000pw}
F.~Hautmann and D.~E.~Soper,
%``Color transparency in deeply inelastic diffraction,''
Phys. Rev. D \textbf{63}, 011501 (2001)
doi:10.1103/PhysRevD.63.011501
[arXiv:hep-ph/0008224 [hep-ph]].
%51 citations counted in INSPIRE as of 29 Feb 2024

%\cite{Beuf:2024msh}
\bibitem{Beuf:2024msh}
G.~Beuf, T.~Lappi, H.~M\"antysaari, R.~Paatelainen and J.~Penttala,
%``Diffractive deep inelastic scattering at NLO in the dipole picture,''
[arXiv:2401.17251 [hep-ph]].
%1 citations counted in INSPIRE as of 19 Mar 2024

%\cite{Trentadue:1993ka}
\bibitem{Trentadue:1993ka}
L.~Trentadue and G.~Veneziano,
%``Fracture functions: An Improved description of inclusive hard processes in QCD,''
Phys. Lett. B \textbf{323}, 201-211 (1994)
doi:10.1016/0370-2693(94)90292-5
%260 citations counted in INSPIRE as of 29 Feb 2024

%\cite{Grazzini:1997ih}
\bibitem{Grazzini:1997ih}
M.~Grazzini, L.~Trentadue and G.~Veneziano,
%``Fracture functions from cut vertices,''
Nucl. Phys. B \textbf{519}, 394-404 (1998)
doi:10.1016/S0550-3213(97)00840-7
[arXiv:hep-ph/9709452 [hep-ph]].
%85 citations counted in INSPIRE as of 29 Feb 2024

%\cite{Graudenz:1994dq}
\bibitem{Graudenz:1994dq}
D.~Graudenz,
%``One particle inclusive processes in deeply inelastic lepton - nucleon scattering,''
Nucl. Phys. B \textbf{432}, 351-376 (1994)
doi:10.1016/0550-3213(94)90606-8
[arXiv:hep-ph/9406274 [hep-ph]].
%98 citations counted in INSPIRE as of 29 Feb 2024

%\cite{deFlorian:1995fd}
\bibitem{deFlorian:1995fd}
D.~de Florian, C.~A.~Garcia Canal and R.~Sassot,
%``Factorization in semiinclusive polarized deep inelastic scattering,''
Nucl. Phys. B \textbf{470}, 195-210 (1996)
doi:10.1016/0550-3213(96)00159-9
[arXiv:hep-ph/9510262 [hep-ph]].
%61 citations counted in INSPIRE as of 29 Feb 2024

%\cite{Collins:1997sr}
\bibitem{Collins:1997sr}
J.~C.~Collins,
%``Proof of factorization for diffractive hard scattering,''
Phys. Rev. D \textbf{57}, 3051-3056 (1998)
[erratum: Phys. Rev. D \textbf{61}, 019902 (2000)]
doi:10.1103/PhysRevD.61.019902
[arXiv:hep-ph/9709499 [hep-ph]].
%496 citations counted in INSPIRE as of 27 Mar 2024

%\cite{Anselmino:2011ss}
\bibitem{Anselmino:2011ss}
M.~Anselmino, V.~Barone and A.~Kotzinian,
%``SIDIS in the target fragmentation region: Polarized and transverse momentum dependent fracture functions,''
Phys. Lett. B \textbf{699}, 108-118 (2011)
doi:10.1016/j.physletb.2011.03.067
[arXiv:1102.4214 [hep-ph]].
%31 citations counted in INSPIRE as of 29 Feb 2024

%\cite{Anselmino:2011bb}
\bibitem{Anselmino:2011bb}
M.~Anselmino, V.~Barone and A.~Kotzinian,
%``Double hadron lepto-production in the current and target fragmentation regions,''
Phys. Lett. B \textbf{706}, 46-52 (2011)
doi:10.1016/j.physletb.2011.10.064
[arXiv:1109.1132 [hep-ph]].
%17 citations counted in INSPIRE as of 29 Feb 2024

%\cite{Boglione:2016bph}
\bibitem{Boglione:2016bph}
M.~Boglione, J.~Collins, L.~Gamberg, J.~O.~Gonzalez-Hernandez, T.~C.~Rogers and N.~Sato,
%``Kinematics of Current Region Fragmentation in Semi-Inclusive Deeply Inelastic Scattering,''
Phys. Lett. B \textbf{766}, 245-253 (2017)
doi:10.1016/j.physletb.2017.01.021
[arXiv:1611.10329 [hep-ph]].
%43 citations counted in INSPIRE as of 29 Feb 2024

%\cite{Chen:2021vby}
\bibitem{Chen:2021vby}
K.~B.~Chen, J.~P.~Ma and X.~B.~Tong,
%``Matching of fracture functions for SIDIS in target fragmentation region,''
JHEP \textbf{11}, 038 (2021)
doi:10.1007/JHEP11(2021)038
[arXiv:2108.13582 [hep-ph]].
%3 citations counted in INSPIRE as of 19 Mar 2024

%\cite{Chen:2023wsi}
\bibitem{Chen:2023wsi}
K.~B.~Chen, J.~P.~Ma and X.~B.~Tong,
%``Twist-3 contributions in semi-inclusive DIS in the target fragmentation region,''
Phys. Rev. D \textbf{108}, no.9, 9 (2023)
doi:10.1103/PhysRevD.108.094015
[arXiv:2308.11251 [hep-ph]].
%2 citations counted in INSPIRE as of 25 Mar 2024

%\cite{Chen:2024brp}
\bibitem{Chen:2024brp}
K.~B.~Chen, J.~P.~Ma and X.~B.~Tong,
%``Gluonic contributions to semi-inclusive DIS in the target fragmentation region,''
[arXiv:2402.15112 [hep-ph]].
%0 citations counted in INSPIRE as of 25 Mar 2024

%\cite{Guo:2023uis}
\bibitem{Guo:2023uis}
Y.~Guo and F.~Yuan,
%``Explore the Nucleon Tomography through Di-hadron Correlation in Opposite Hemisphere in Deep Inelastic Scattering,''
[arXiv:2312.01008 [hep-ph]].
%2 citations counted in INSPIRE as of 20 Mar 2024

%\cite{CLAS:2022sqt}
\bibitem{CLAS:2022sqt}
H.~Avakian \textit{et al.} [CLAS],
%``Observation of Correlations between Spin and Transverse Momenta in Back-to-Back Dihadron Production at CLAS12,''
Phys. Rev. Lett. \textbf{130}, no.2, 022501 (2023)
doi:10.1103/PhysRevLett.130.022501
[arXiv:2208.05086 [hep-ex]].
%7 citations counted in INSPIRE as of 13 Mar 2024

%\cite{Boer:2011fh}
\bibitem{Boer:2011fh}
D.~Boer, M.~Diehl, R.~Milner, R.~Venugopalan, W.~Vogelsang, D.~Kaplan, H.~Montgomery, S.~Vigdor, A.~Accardi and E.~C.~Aschenauer, \textit{et al.}
%``Gluons and the quark sea at high energies: Distributions, polarization, tomography,''
[arXiv:1108.1713 [nucl-th]].
%687 citations counted in INSPIRE as of 20 Mar 2024

%\cite{AbelleiraFernandez:2012cc}
\bibitem{AbelleiraFernandez:2012cc}
J.~L.~Abelleira Fernandez \textit{et al.} [LHeC Study Group],
%``A Large Hadron Electron Collider at CERN: Report on the Physics and Design Concepts for Machine and Detector,''
J. Phys. G \textbf{39}, 075001 (2012)
doi:10.1088/0954-3899/39/7/075001
[arXiv:1206.2913 [physics.acc-ph]].
%811 citations counted in INSPIRE as of 25 Mar 2024

%\cite{Accardi:2012qut}
\bibitem{Accardi:2012qut}
A.~Accardi, J.~L.~Albacete, M.~Anselmino, N.~Armesto, E.~C.~Aschenauer, A.~Bacchetta, D.~Boer, W.~K.~Brooks, T.~Burton and N.~B.~Chang, \textit{et al.}
%``Electron Ion Collider: The Next QCD Frontier: Understanding the glue that binds us all,''
Eur. Phys. J. A \textbf{52}, no.9, 268 (2016)
doi:10.1140/epja/i2016-16268-9
[arXiv:1212.1701 [nucl-ex]].
%1584 citations counted in INSPIRE as of 26 Mar 2024

%\cite{AbdulKhalek:2021gbh}
\bibitem{AbdulKhalek:2021gbh}
R.~Abdul Khalek, A.~Accardi, J.~Adam, D.~Adamiak, W.~Akers, M.~Albaladejo, A.~Al-bataineh, M.~G.~Alexeev, F.~Ameli and P.~Antonioli, \textit{et al.}
%``Science Requirements and Detector Concepts for the Electron-Ion Collider: EIC Yellow Report,''
Nucl. Phys. A \textbf{1026}, 122447 (2022)
doi:10.1016/j.nuclphysa.2022.122447
[arXiv:2103.05419 [physics.ins-det]].
%780 citations counted in INSPIRE as of 28 Mar 2024

%\cite{Gross:2022hyw}
\bibitem{Gross:2022hyw}
F.~Gross, E.~Klempt, S.~J.~Brodsky, A.~J.~Buras, V.~D.~Burkert, G.~Heinrich, K.~Jakobs, C.~A.~Meyer, K.~Orginos and M.~Strickland, \textit{et al.}
%``50 Years of Quantum Chromodynamics,''
Eur. Phys. J. C \textbf{83}, 1125 (2023)
doi:10.1140/epjc/s10052-023-11949-2
[arXiv:2212.11107 [hep-ph]].
%63 citations counted in INSPIRE as of 28 Mar 2024

%\cite{Achenbach:2023pba}
\bibitem{Achenbach:2023pba}
P.~Achenbach, D.~Adhikari, A.~Afanasev, F.~Afzal, C.~A.~Aidala, A.~Al-bataineh, D.~K.~Almaalol, M.~Amaryan, D.~Androic and W.~R.~Armstrong, \textit{et al.}
%``The Present and Future of QCD,''
[arXiv:2303.02579 [hep-ph]].
%34 citations counted in INSPIRE as of 27 Mar 2024

%\cite{Mulders:1995dh}
\bibitem{Mulders:1995dh}
P.~J.~Mulders and R.~D.~Tangerman,
%``The Complete tree level result up to order 1/Q for polarized deep inelastic leptoproduction,''
Nucl. Phys. B \textbf{461}, 197-237 (1996)
[erratum: Nucl. Phys. B \textbf{484}, 538-540 (1997)]
doi:10.1016/0550-3213(95)00632-X
[arXiv:hep-ph/9510301 [hep-ph]].
%983 citations counted in INSPIRE as of 20 Mar 2024

%\cite{Boer:1997nt}
\bibitem{Boer:1997nt}
D.~Boer and P.~J.~Mulders,
%``Time reversal odd distribution functions in leptoproduction,''
Phys. Rev. D \textbf{57}, 5780-5786 (1998)
doi:10.1103/PhysRevD.57.5780
[arXiv:hep-ph/9711485 [hep-ph]].
%986 citations counted in INSPIRE as of 28 Mar 2024

%\cite{Bacchetta:2006tn}
\bibitem{Bacchetta:2006tn}
A.~Bacchetta, M.~Diehl, K.~Goeke, A.~Metz, P.~J.~Mulders and M.~Schlegel,
%``Semi-inclusive deep inelastic scattering at small transverse momentum,''
JHEP \textbf{02}, 093 (2007)
doi:10.1088/1126-6708/2007/02/093
[arXiv:hep-ph/0611265 [hep-ph]].
%771 citations counted in INSPIRE as of 27 Mar 2024

%\cite{Collins:2002kn}
\bibitem{Collins:2002kn}
J.~C.~Collins,
%``Leading twist single transverse-spin asymmetries: Drell-Yan and deep inelastic scattering,''
Phys. Lett. B \textbf{536}, 43-48 (2002)
doi:10.1016/S0370-2693(02)01819-1
[arXiv:hep-ph/0204004 [hep-ph]].
%1011 citations counted in INSPIRE as of 28 Mar 2024

%\cite{Ji:2002aa}
\bibitem{Ji:2002aa}
X.~d.~Ji and F.~Yuan,
%``Parton distributions in light cone gauge: Where are the final state interactions?,''
Phys. Lett. B \textbf{543}, 66-72 (2002)
doi:10.1016/S0370-2693(02)02384-5
[arXiv:hep-ph/0206057 [hep-ph]].
%474 citations counted in INSPIRE as of 20 Mar 2024

%\cite{Belitsky:2002sm}
\bibitem{Belitsky:2002sm}
A.~V.~Belitsky, X.~Ji and F.~Yuan,
%``Final state interactions and gauge invariant parton distributions,''
Nucl. Phys. B \textbf{656}, 165-198 (2003)
doi:10.1016/S0550-3213(03)00121-4
[arXiv:hep-ph/0208038 [hep-ph]].
%684 citations counted in INSPIRE as of 20 Mar 2024

%\cite{DeTar:1974vx}
\bibitem{DeTar:1974vx}
C.~E.~DeTar, S.~D.~Ellis and P.~V.~Landshoff,
%``Final State Interactions in Large Transverse Momentum Lepton and Hadron Production,''
Nucl. Phys. B \textbf{87}, 176-188 (1975)
doi:10.1016/0550-3213(75)90260-6
%112 citations counted in INSPIRE as of 29 Feb 2024

%\cite{Cardy:1974vq}
\bibitem{Cardy:1974vq}
J.~L.~Cardy and G.~A.~Winbow,
%``The Absence of Final State Interaction Corrections to the Drell-Yan Formula for Massive Lepton Pair Production,''
Phys. Lett. B \textbf{52}, 95-96 (1974)
doi:10.1016/0370-2693(74)90729-1
%101 citations counted in INSPIRE as of 29 Feb 2024

%\cite{Collins:1992cv}
\bibitem{Collins:1992cv}
J.~C.~Collins, L.~Frankfurt and M.~Strikman,
%``Diffractive hard scattering with a coherent pomeron,''
Phys. Lett. B \textbf{307}, 161-168 (1993)
doi:10.1016/0370-2693(93)90206-W
[arXiv:hep-ph/9212212 [hep-ph]].
%163 citations counted in INSPIRE as of 29 Feb 2024

%\cite{Collins:2001ga}
\bibitem{Collins:2001ga}
J.~C.~Collins,
%``Factorization in hard diffraction,''
J. Phys. G \textbf{28}, 1069-1078 (2002)
doi:10.1088/0954-3899/28/5/327
[arXiv:hep-ph/0107252 [hep-ph]].
%42 citations counted in INSPIRE as of 19 Mar 2024

%\cite{CDF:2000rua}
\bibitem{CDF:2000rua}
T.~Affolder \textit{et al.} [CDF],
%``Diffractive Dijets with a Leading Antiproton in $\bar{p}p$ Collisions at $\sqrt{s} = 1800$ GeV,''
Phys. Rev. Lett. \textbf{84}, 5043-5048 (2000)
doi:10.1103/PhysRevLett.84.5043
%270 citations counted in INSPIRE as of 29 Feb 2024

%\cite{Pasechnik:2023mdd}
\bibitem{Pasechnik:2023mdd}
R.~Pasechnik and M.~Ta\v{s}evsk\'y,
%``Multi-dimensional hadron structure through the lens of gluon Wigner distribution,''
[arXiv:2310.10793 [hep-ph]].
%4 citations counted in INSPIRE as of 26 Mar 2024

%\cite{Hagiwara:2016kam}
\bibitem{Hagiwara:2016kam}
Y.~Hagiwara, Y.~Hatta and T.~Ueda,
%``Wigner, Husimi, and generalized transverse momentum dependent distributions in the color glass condensate,''
Phys. Rev. D \textbf{94}, no.9, 094036 (2016)
doi:10.1103/PhysRevD.94.094036
[arXiv:1609.05773 [hep-ph]].
%52 citations counted in INSPIRE as of 25 Mar 2024

%\cite{Yao:2018vcg}
\bibitem{Yao:2018vcg}
X.~Yao, Y.~Hagiwara and Y.~Hatta,
%``Computing the gluon Sivers function at small-$x$,''
Phys. Lett. B \textbf{790}, 361-366 (2019)
doi:10.1016/j.physletb.2019.01.029
[arXiv:1812.03959 [hep-ph]].
%21 citations counted in INSPIRE as of 12 Mar 2024

%\cite{Iancu:2017fzn}
\bibitem{Iancu:2017fzn}
E.~Iancu and A.~H.~Rezaeian,
%``Elliptic flow from color-dipole orientation in pp and pA collisions,''
Phys. Rev. D \textbf{95}, no.9, 094003 (2017)
doi:10.1103/PhysRevD.95.094003
[arXiv:1702.03943 [hep-ph]].
%53 citations counted in INSPIRE as of 25 Mar 2024

%\cite{Linek:2023kga}
\bibitem{Linek:2023kga}
B.~Linek, A.~\L{}uszczak, M.~\L{}uszczak, R.~Pasechnik, W.~Sch\"afer and A.~Szczurek,
%``Probing proton structure with $ \textrm{c}\overline{\textrm{c}} $ correlations in ultraperipheral pA collisions,''
JHEP \textbf{10}, 179 (2023)
doi:10.1007/JHEP10(2023)179
[arXiv:2308.00457 [hep-ph]].
%3 citations counted in INSPIRE as of 28 Mar 2024

%\cite{Kovchegov:2003dm}
\bibitem{Kovchegov:2003dm}
Y.~V.~Kovchegov, L.~Szymanowski and S.~Wallon,
%``Perturbative odderon in the dipole model,''
Phys. Lett. B \textbf{586}, 267-281 (2004)
doi:10.1016/j.physletb.2004.02.036
[arXiv:hep-ph/0309281 [hep-ph]].
%102 citations counted in INSPIRE as of 01 Mar 2024

%\cite{Hatta:2005as}
\bibitem{Hatta:2005as}
Y.~Hatta, E.~Iancu, K.~Itakura and L.~McLerran,
%``Odderon in the color glass condensate,''
Nucl. Phys. A \textbf{760}, 172-207 (2005)
doi:10.1016/j.nuclphysa.2005.05.163
[arXiv:hep-ph/0501171 [hep-ph]].
%124 citations counted in INSPIRE as of 01 Mar 2024

%\cite{Dumitru:2018vpr}
\bibitem{Dumitru:2018vpr}
A.~Dumitru, G.~A.~Miller and R.~Venugopalan,
%``Extracting many-body color charge correlators in the proton from exclusive DIS at large Bjorken x,''
Phys. Rev. D \textbf{98}, no.9, 094004 (2018)
doi:10.1103/PhysRevD.98.094004
[arXiv:1808.02501 [hep-ph]].
%29 citations counted in INSPIRE as of 01 Mar 2024

%\cite{Benic:2023ybl}
\bibitem{Benic:2023ybl}
S.~Beni\'c, D.~Horvati\'c, A.~Kaushik and E.~A.~Vivoda,
%``Exclusive \ensuremath{\eta}c production from small-x evolved Odderon at an electron-ion collider,''
Phys. Rev. D \textbf{108}, no.7, 074005 (2023)
doi:10.1103/PhysRevD.108.074005
[arXiv:2306.10626 [hep-ph]].
%1 citations counted in INSPIRE as of 26 Mar 2024

%\cite{Zhou:2013gsa}
\bibitem{Zhou:2013gsa}
J.~Zhou,
%``Transverse single spin asymmetries at small x and the anomalous magnetic moment,''
Phys. Rev. D \textbf{89}, no.7, 074050 (2014)
doi:10.1103/PhysRevD.89.074050
[arXiv:1308.5912 [hep-ph]].
%55 citations counted in INSPIRE as of 01 Mar 2024

%\cite{Boer:2015pni}
\bibitem{Boer:2015pni}
D.~Boer, M.~G.~Echevarria, P.~Mulders and J.~Zhou,
%``Single spin asymmetries from a single Wilson loop,''
Phys. Rev. Lett. \textbf{116}, no.12, 122001 (2016)
doi:10.1103/PhysRevLett.116.122001
[arXiv:1511.03485 [hep-ph]].
%61 citations counted in INSPIRE as of 29 Feb 2024

%\cite{Boussarie:2019vmk}
\bibitem{Boussarie:2019vmk}
R.~Boussarie, Y.~Hatta, L.~Szymanowski and S.~Wallon,
%``Probing the Gluon Sivers Function with an Unpolarized Target: GTMD Distributions and the Odderons,''
Phys. Rev. Lett. \textbf{124}, no.17, 172501 (2020)
doi:10.1103/PhysRevLett.124.172501
[arXiv:1912.08182 [hep-ph]].
%29 citations counted in INSPIRE as of 01 Mar 2024

%\cite{Metz:2011wb}
\bibitem{Metz:2011wb}
A.~Metz and J.~Zhou,
%``Distribution of linearly polarized gluons inside a large nucleus,''
Phys. Rev. D \textbf{84}, 051503 (2011)
doi:10.1103/PhysRevD.84.051503
[arXiv:1105.1991 [hep-ph]].
%150 citations counted in INSPIRE as of 29 Feb 2024
\end{thebibliography}
\end{document}